\documentclass[twocolumn, prb, showpacs, preprintnumbers,amsmath,amssymb]{revtex4-1}

\usepackage{graphicx}
\usepackage{dcolumn}
\usepackage{bm}

\newcommand{\beq}{\begin{equation}}
\newcommand{\eeq}{\end{equation}}
\newcommand{\bdis}{\begin{displaymath}}
\newcommand{\edis}{\end{displaymath}}
\newcommand{\bea}{\begin{eqnarray}}
\newcommand{\eea}{\end{eqnarray}}
\newcommand{\barr}{\begin{array}}
\newcommand{\earr}{\end{array}}
\newcommand{\bfig}{\begin{figure}[!]}
\newcommand{\efig}{\end{figure}}

\begin{document}

\preprint{}

\title{Non-equilibrium thermodynamics of the spin Seebeck and spin Peltier effects}

\author{Vittorio Basso, Elena Ferraro, Alessandro Magni, Alessandro Sola, Michaela Kuepferling and Massimo Pasquale}
\affiliation{Istituto Nazionale di Ricerca Metrologica, Strada delle Cacce 91, 10135 Torino, Italy}

\date{\today}

\begin{abstract}
We study the problem of magnetization and heat currents and their associated thermodynamic forces in a magnetic system by focusing on the magnetization transport in ferromagnetic insulators like YIG. The resulting theory is applied to the longitudinal spin Seebeck and spin Peltier effects. By focusing on the specific geometry with one YIG layer and one Pt layer, we obtain the optimal conditions for generating large magnetization currents into Pt or large temperature effects in YIG. The theoretical predictions are compared with experiments from the literature permitting to derive the values of the thermomagnetic coefficients of YIG: the magnetization diffusion length $l_M \sim 0.4 \, \mu$m and the absolute thermomagnetic power coefficient $\epsilon_M \sim 10^{-2}$ TK$^{-1}$.
\end{abstract}

\maketitle

\section{Introduction}

The recent discovery of the longitudinal spin Seebeck effect in ferromagnetic insulators has raised a renewed interest in the non equilibrium thermodynamics of spin or magnetization currents \cite{Uchida-2010}. Experiments have shown that a temperature gradient applied across an electrically insulating magnetic material is able to inject a spin current into an adjacent metal, where the spin polarization is revealed by means of the inverse spin Hall effect (ISHE) \cite{Siegel-2014, Uchida-2014}. Typical experiments have been performed by using ferrimagnets, like the yttrium iron garnet (Y$_3$Fe$_5$O$_{12}$, YIG) as insulating magnetic material and Pt or other noble metals, as conductors \cite{Siegel-2014, Uchida-2014}. In analogy to thermoelectrics, the reciprocal of the spin Seebeck effect has been called spin Peltier effect \cite{Flipse-2014}. This reciprocal effect has been recently observed by using the spin Hall effect of Pt as spin current injector and observing the thermal effects on YIG \cite{Flipse-2014}. All these experiments show that the magnetization current can propagate along different media using different type of carriers. While spin currents in metals are associated to the unbalance in the spin polarization of conduction electrons, in magnetic insulators the magnetization transport is due to spin waves or magnons \cite{Adachi-2013}. Spin Seebeck and spin Peltier experiments reveal that the magnetization current carried by magnons in the magnetic insulator can be transformed into a spin current carried by electrons and viceversa. The mechanism of this conversion is seen as the interfacial s-d coupling between the localized magnetic moment of the ferromagnet (which is often due to d shell electrons) and the conduction electrons of the metal (which are often s shell electrons) \cite{Uchida-2013, Hoffmann-2013, Zhang-2012}. 

The thermodynamics of thermo-magneto-electric effects, i.e. spin caloritronics, has been already developed for metals by adding the spin degree of freedom to the thermo-electricity theory \cite{Johnson-1987,Bauer-2012}. However, spin caloritronics cannot be directly applied to electrical insulating magnetic materials like YIG. Therefore it is necessary to develop a more general theory which could be applied to both conductors and insulators. The formulations of the problem present in the literature often focus on the microscopic origin without paying much attention to the formal thermodynamic theory that is expected as a result. Refs.\cite{Xiao-2010, Xiao-2010b, Adachi-2013, Schreier-2013, Chotorlishvili-2015} describe the non equilibrium magnon distribution through an effective magnon temperature different from the lattice temperature. However from an experimental point of view in Ref.\cite{Agrawal-2013} it was observed a close correspondence between the spatial dependencies of the exchange magnon and phonon temperatures. The Boltzmann approach for magnon transport was used in Ref.\cite{Tulapurkar-2011, Zhang-2012, Rezende-2014, Rezende-2014b, Rezende-2016}, combined by a YIG/Pt interface coupling \cite{Zhang-2012, Hoffmann-2013}. Within these approaches the spin accumulation and the magnon accumulation take the role of an effective force able to drive the magnetization current. The use of different quantities between the two sides of a junction requires therefore the introduction of a spin convertance to account for the magnon current induced by spin accumulation and the spin current created by magnon accumulation \cite{Zhang-2012b}.

The aim of the present paper is to define the macroscopic non-equilibrium thermodynamics picture for the problems related to magnetization currents that could be used independently of the specific magnetic moment carrier. To this aim we start from the results of the thermodynamic theory of Johnson and Silsbee \cite{Johnson-1987}. The main difference with respect to the classical theories of the thermoelectric effects is that the magnetization current density $j_M$ is not continuous. The magnetic moment can both flow through a magnetization current but also can be locally absorbed and generated by sinks and sources. Here, by limiting the analysis to the scalar case, we state the simplest possible continuity equation for the magnetization. As a result we find that the potential for the magnetization current is the difference $H^* = H-H_{eq}$ between the magnetic field $H$ and the equilibrium field $H_{eq}$. The gradient of the potential $\nabla H^*$ is the thermodynamic force to be associated to the magnetization current.

With this definition it is then possible to state the constitutive equations for the joint magnetization and heat transport and to identify the absolute thermomagnetic power coefficient $\epsilon_M$ relating the gradient of the potential of the magnetization current $\mu_0\nabla H^*$ with the temperature gradient $\nabla T$, in analogy with thermoelectricity. The same coefficient also determines the spin Peltier heat current $\epsilon_M T j_M$ when the system is subjected to a magnetization current.

In the present work we apply the previous arguments to describe the generation of a magnetization current by the spin Seebeck effect and the heat transport caused by the spin Peltier effect. To this end we have to complement the constitutive equations for the thermo-magnetic active material (YIG) with the equations for the spin Hall active layer (Pt). Once the equations for the two materials are written by using the same thermodynamic formalism, one can apply the theory to solve specific problems of magnetization current traversing different layers. The diffusion length for the magnetization current $l_M = (\mu_0 \sigma_M \tau_M)^{1/2}$ is related to intrinsic properties of each material: the magnetization conductivity $\sigma_M$ and the time constant $\tau_M$, describing how fast the system is able to absorb the magnetic moment in excess. We are also able to show that the passage of the magnetization current from one layer to the other is governed by the ratio between $l_M/\tau_M$ of the two layers.

By focusing on the specific geometry with one YIG layer and one Pt layer, we obtain the optimal conditions for generating large magnetization currents into Pt in the case of the spin Seebeck effect and for generating large heat current in YIG in the case of spin Peltier effects. In both cases we find that efficient injection is obtained when the thickness of the injecting layer is larger than the critical thickness $l_M$ as recently experiments confirm \cite{Kehlberger-2015}. We finally determine the values of the thermomagnetic coefficients of YIG by comparing the theory to recent experiments \cite{Sola-2015, Flipse-2014}.

The paper is organized as follows. In Section II we first discuss the thermodynamic properties of an out-of-equilibrium but spatially uniform magnetic system \cite{Bertotti-2006} and on that basis we introduce, for non spatially uniform system, the currents and the thermodynamic forces in analogy with the non equilibrium thermodynamics of thermoelectric effects \cite{Callen-1985}. In Section III we set the constitutive equations for the magnetization and heat transport in both an insulating ferrimagnet and a metal with the spin Hall effect. Section IV is devoted to the solutions of the magnetization current problem. In Section V we focus on the specific longitudinal spin Seebeck geometry and on the spin Peltier effect. Finally some conclusive remarks are drawn in Section VI.

\section{Thermodynamics of magnetization currents}

\subsection{Thermodynamics of uniform magnetic systems}

\begin{figure}[htb]
\narrowtext 
\centering
\includegraphics[width=6cm]{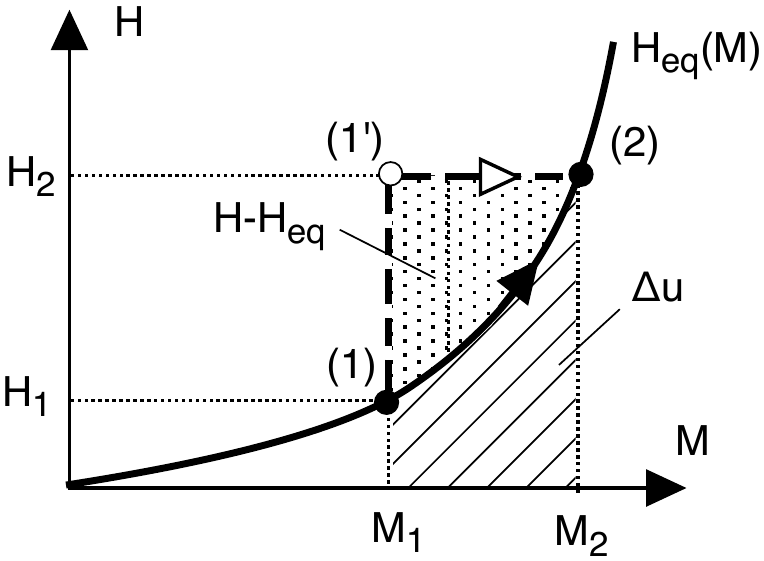}
\caption{Equilibrium path (solid line) and out-of-equilibrium path (dashed line) connecting the equilibrium states (1) and (2) in the $H$ versus $M$ diagram of a magnetic material. $H_{eq}(M)$ is the equilibrium state equation at constant entropy. (1$^{\prime}$) is an out-of-equilibrium state obtained by the sudden change of the field from $H_1$ to $H_2$. In the relaxation path from (1$^{\prime}$) to (2) the work is $\mu_0HdM$, the internal energy change is $du=\mu_0H_{eq}dM$ and the entropy production is $T\sigma_sdt = \mu_0(H-H_{eq})dM$. The relaxation equation is Eq.(\ref{EQ:dMdt_hom})} \label{FIG:1}
\end{figure}

We consider a magnetic system that can be described by a scalar magnetization $M$. Suitable systems can be ferromagnetic or ferrimagnetic materials where an easy axis is present, due for example to an anisotropic crystal structure, along which all the vector quantities are lying. We take spatially uniform quantities and all extensive quantities as volume densities. The derivative of the internal energy density $u(s,M)$ with respect to the magnetization at constant entropy density $s$, gives the equilibrium state equation

\beq
H_{eq} = \frac{1}{\mu_0}\frac{\partial u}{\partial M}\biggr\rvert_s
\label{EQ:H_eq}
\eeq

\noindent where $\mu_0$ is the magnetic permeability of vacuum. In equilibrium the magnetic field $H$ is equal to the state equation $H=H_{eq}(M,s)$. When $H$ is different from its equilibrium value $H_{eq}$ the system state will try to reach the equilibrium by the action of dissipative processes. In a generic out-of-equilibrium situation the variation of internal energy must take into account that dissipative processes correspond to an entropy production. The energy balance then reads

\beq
du = T ds + \mu_0H dM - T \sigma_s dt
\label{EQ:du}
\eeq

\noindent where $T = \partial u/\partial s$ is the temperature, $\mu_0HdM$ is the infinitesimal work done on the system, $\sigma_s$ is the entropy production rate, which has to be a definite positive term and $t$ is time. When approaching equilibrium, the magnetization $M$ will change until the equilibrium condition $H=H_{eq}(M)$ will be reached. The typical situation is sketched in Fig.\ref{FIG:1} showing two processes connecting the equilibrium states (1) and (2). The equilibrium path (solid line) corresponds to the slow variation of field $H$ from $H_1$ to  $H_2$ through the equilibrium state equation $H=H_{eq}(M)$. The out-of-equilibrium path (dashed line) passes  through the out-of-equilibrium state (1$^{\prime}$) and corresponds to the sudden variation of the field from $H_1$ to $H_2$ and to the subsequent time relaxation. As the initial and final states are always equilibrium states, the final internal energy variation must be the same for any process. This is obtained by assuming that the part of the work going into the internal energy is always the equilibrium one. Then, by inserting $du=\mu_0H_{eq}dM$ (from Eq.(\ref{EQ:H_eq})) into Eq.(\ref{EQ:du}) at constant entropy ($ds=0$) we find
the expression for the entropy production rate

\beq
\sigma_s = \mu_0 \frac{H-H_{eq}}{T}\frac{dM}{dt}.
\label{EQ:sigma_s_hom}
\eeq

\noindent As expected, the entropy production rate is the product of a generalized force, or affinity, represented by the term $\mu_0 (H-H_{eq})/T$, times a generalized flux, or velocity, represented by $dM/dt$ \cite{Callen-1985}. If the distance from equilibrium is not too large one can consider the linear system approximation and assume the velocity to be proportional to the affinity. It is appropriate to describe this fact by introducing a typical time constant $\tau_M$ for the process by defining 

\beq
\frac{dM}{dt} = \frac{H-H_{eq}}{\tau_M},
\label{EQ:dMdt_hom}
\eeq

\noindent where the temperature $T$ and $\mu_0$ appearing in the generalized force, have been incorporated into the definition of the time constant. Eq.(\ref{EQ:dMdt_hom}) provides a kinetic equation for the magnetization describing the time relaxation from a generic out-of-equilibrium state by showing that the velocity $dM/dt$ depends on the distance from equilibrium $H-H_{eq}$ (see Fig.1). 

The interesting physics behind Eq.(\ref{EQ:dMdt_hom}) is that it also expresses the non conservation of the magnetic moment with the presence of sources and sinks, although the total angular momentum for an isolated system is conserved. As a matter of fact in the solid state there is a huge reservoir of angular momentum available (electrons, nuclei, etc) and only a very small part of it is associated to the magnetic moment. As a result, the magnetization can be easily varied by exchanging angular momentum with the reservoir constituted by non magnetic degrees of freedoms. With this in mind, the physical meaning of Eq.(\ref{EQ:dMdt_hom}) is to express how fast the angular momentum from the magnetization subsystem can be exchanged with the reservoir.

Finally, as it happens in many problems involving a non conserved magnetization, also the internal energy is a non conserved quantity. To avoid the problem, we pass to the enthalpy potential $u_e = u -\mu_0HM$ which contains the magnetic field $H$ and the entropy $s$ as independent variables. Dealing with out-of-equilibrium processes, the potential $u_e$ is also a non equilibrium one which depends on the magnetization $M$ as an internal variable. From Eq.(\ref{EQ:du}), the enthalpy variation is

\beq
du_e = Tds - \mu_0MdH  - \mu_0(H-H_{eq}) dM
\label{EQ:du_e}
\eeq

\noindent where we have used the definition of the entropy production of Eq.(\ref{EQ:sigma_s_hom}). The expression for the variation of the enthalpy potential (\ref{EQ:du_e}), together with the kinetic equation (\ref{EQ:dMdt_hom}), constitutes the out-of-equilibrium thermodynamics of the system \cite{Bertotti-2006} and can be employed to build up the thermodynamics of fluxes and forces.

\subsection{Thermodynamics of fluxes and forces}

We now pass from the out-of-equilibrium thermodynamics of a spatially uniform magnetic system to the problem of having a non uniform situation involving currents of the extensive variables, entropy and magnetization, and the associated thermodynamic forces \cite{Callen-1985}. Both the extensive and intensive variable are now allowed to vary as a function of space coordinates $\boldsymbol{r}$. In the case of extensive variables the volume densities are intended as moving averages over a small volume $\Delta V$ around the point $\boldsymbol{r}$. As the magnetization is a non conserved quantity, we need to explicitly express the fact that any magnetization change $dM$ is in part drawn from the reservoir of angular momentum, which is external to the thermodynamic system, and in part exchanged between the surrounding regions of the thermodynamic system itself, giving rise to a current of magnetic moment $\boldsymbol{j}_{M}$. The sources and sinks of the magnetic moment are exactly those described in the previous Section by Eq.(\ref{EQ:dMdt_hom}), then we can immediately write a continuity equation for the magnetization by extending Eq.(\ref{EQ:dMdt_hom}), obtaining

\beq
\frac{\partial M}{\partial t} +\nabla \cdot \boldsymbol{j}_{M} = \frac{H - H_{eq}(M)}{\tau_M}.
\label{EQ:dM_dt}
\eeq

\noindent Next, as it is usually done in the non equilibrium theory of fluxes and forces \cite{Callen-1985}, we use Eq.(\ref{EQ:du_e}) to pass to the entropy representation by writing the entropy variation

\beq
ds = \frac{1}{T}du_e + \frac{\mu_0M}{T} dH + \frac{\mu_0(H-H_{eq})}{T} dM.
\label{EQ:ds}
\eeq

\noindent As we aim to define the entropy current as a function of the other currents, we have to look at the previous equation in search for the variations of the extensive variables. Eq.(\ref{EQ:ds}) contains the variation of the enthalpy $du_e$ and the magnetization $dM$ which both have associated currents, while the variation of the magnetic field $dH$ does not corresponds to any current and has not to be taken into account in the definition of the entropy current. Then we define

\beq
\boldsymbol{j}_s = \frac{1}{T}\boldsymbol{j}_{u_e} + \frac{\mu_0(H-H_{eq})}{T} \boldsymbol{j}_{M}
\label{EQ:j_s_def}
\eeq

\noindent where $\boldsymbol{j}_{s}$ is the entropy current and $\boldsymbol{j}_{u_e}$ is the enthalpy current which obeys the following continuity equation

\beq
\frac{\partial u_e}{\partial t}+\nabla\cdot \boldsymbol{j}_{u_e} = -\mu_0M\frac{\partial H}{\partial t},
\label{EQ:j_e_def}
\eeq

\noindent from which one notices that the enthalpy is conserved if the field $H$ is constant in time. The continuity equation for the magnetization is Eq.(\ref{EQ:dM_dt}) and finally the entropy obeys the continuity equation 

\beq
\frac{\partial s}{\partial t}+\nabla \cdot \boldsymbol{j}_s = \sigma_s.
\label{EQ:cont_s}
\eeq

\noindent As it is done in the classical treatment\cite{Callen-1985,deGroot-1984}, one expresses the entropy production rate $\sigma_s$ in terms of a sum of products of each current times its thermodynamic force. By using Eqs.(\ref{EQ:ds})-(\ref{EQ:j_e_def}) into Eq.(\ref{EQ:cont_s}) and introducing the heat current as $\boldsymbol{j}_q=T\boldsymbol{j}_s$, after a few passages one obtains

\beq
\sigma_s=\nabla\left(\frac{1}{T}\right)\cdot\boldsymbol{j}_q+\mathcal{F}_M\cdot \boldsymbol{j}_M+\frac{1}{\tau_M}\frac{\mu_0\left(H-H_{eq}\right)^2}{T}
\label{EQ:sigma_s}
\eeq

\noindent where we have defined the thermodynamic force associated to the magnetization current

\beq
\mathcal{F}_M= \frac{1}{T}\mu_0\nabla\left(H - H_{eq}\right).
\label{EQ:F_M}
\eeq

\noindent In Eq.(\ref{EQ:sigma_s}) we see the products of the heat current $\boldsymbol{j}_q$ times its force $\nabla ({1}/{T})$, of the magnetization current $\boldsymbol{j}_M$ times its force $\mathcal{F}_M$ and the last term which is exactly the entropy production associated with the out-of-equilibrium homogeneous processes and not to the fluxes. The last term can be also recognized as entropy production of Eq.(\ref{EQ:sigma_s_hom}) where the affinity is $\mu_0(H-H_{eq})/T$ and the magnetization change $dM/dt$ is $(H-H_{eq})/\tau_{M}$ as given by Eq.(\ref{EQ:dMdt_hom}).

As a main result we have found that the gradient of the distance from equilibrium Eq.(\ref{EQ:F_M}) is the generalized force associated with the magnetization current $\boldsymbol{j}_M$. For simplicity we define $H^* = H-H_{eq}$ to specify the distance from equilibrium and we observe that the driving force of the magnetization current appears as soon as the system is brought out-of-equilibrium. In that case the system may find more effective to draw magnetization from the surroundings rather than from the local spin reservoir. The strength of this effect is given by a further parameter, the magnetization conductivity $\sigma_M$, which establishes the relationship between the magnetization current $\boldsymbol{j}_M$ and the gradient of $H^*$

\beq
\boldsymbol{j}_M = \sigma_M \mu_0 \nabla H^*.
\label{jM_cost}
\eeq

\noindent $H^*$ can be different from zero in stationary situation every time the material experiences the accumulation of magnetization (i.e. spin accumulation in the case of metallic conductors). We have to notice that even if $H^*$ has the units of a magnetic field, it is not a magnetic field in the sense of the Maxwell equations of electromagnetism. Its status is analogous to the exchange field or the anisotropy field of ferromagnets whose origins is in the quantum mechanics of the solid. $H^*$ represents the thermodynamic reaction of the system for finding itself in an out-of-equilibrium situation. In the following we refer to $H^*$ as the potential for the magnetization current.

\section{Constitutive equations}

Having defined the potential $H^*$ associated with the magnetization current, we are ready to write the constitutive equations for the two materials of interest for the spin Seebeck and spin Peltier effects: a magnetic insulating material with a spin Seebeck effect and a metallic conductor with the spin Hall effect.


\subsection{Thermomagnetic effects in magnetic insulators}
\label{IIIA}
In analogy with the thermoelectric effects \cite{Callen-1985}, we can write the constitutive equation for the joint transport of magnetization and heat by using the potential associated with the magnetization current derived in the previous Section. The general case which includes the presence of electric current is reported in Appendix A. Here we limit to insulators and we take currents and forces in one dimension ($\nabla_x = \partial/\partial x$). The equations for the thermomagnetic effect reads

\begin{align}
\label{EQ:jM_const}
j_M & = \sigma_M \, \mu_0 \nabla_x H^* - \sigma_M \epsilon_M  \, \nabla_x T \\
j_q & = \epsilon_M \sigma_M T \mu_0 \nabla_x H^* - (\kappa + \epsilon_M^2 \sigma_M T) \nabla_x T
\label{EQ:jq_const}
\end{align}

\noindent where $\sigma_M$ is the spin conductivity, $\epsilon_M$ is absolute thermomagnetic power coefficient, $j_q$ is the heat current density and $\kappa$ is the thermal conductivity under zero magnetization current. Since the magnetization is not conserved, the magnetization current is not continuous and we have always to add the continuity equation (\ref{EQ:dM_dt}). In non-equilibrium stationary states we always ask the condition $\partial M/\partial t=0$ to be true, so Eq.(\ref{EQ:dM_dt}) becomes

\beq
\nabla_x \, j_{M} = \frac{H^*}{\tau_M}. 
\label{EQ:cont_M}
\eeq

\subsubsection{Uniform temperature gradient}
\label{IIIA1}

If we disregard for the moment the heat currents, the solution of magnetization current problems will correspond to find solutions to the system composed by Eqs.(\ref{EQ:jM_const}) and (\ref{EQ:cont_M}). Under a uniform temperature gradient, where $\nabla T$ is a constant, the second term at the right hand side of Eq.(\ref{EQ:jM_const}) is just a magnetization current density source $j_{MS} = - \sigma_M \epsilon_M  \, \nabla_x T$. Then the solution of 

\beq
j_M = j_{MS} + \sigma_M \, \mu_0 \nabla_x H^*
\label{EQ:jM_isoT}
\eeq

\noindent together with Eq.(\ref{EQ:cont_M}), considering constant coefficients, leads to a differential equation for the potential

\beq
l_M^2 \nabla_x^2 H^* = H^*
\label{EQ:diff_eq}
\eeq

\noindent where

\beq
l_M = (\mu_0 \sigma_M \tau_M)^{1/2}
\label{EQ:lM}
\eeq

\noindent is a material dependent diffusion length. The differential equation (\ref{EQ:diff_eq}) has general solutions in the form

\beq
H^*(x) = H^*_{-} \exp(-x/l_M) + H^*_{+} \exp(x/l_M)
\label{EQ:h_ast}
\eeq

\noindent where $H^*_{-}$ and $H^*_{+}$ are coefficients to be determined on the base of the boundary conditions. By looking at Eqs.(\ref{EQ:jM_const}) and (\ref{EQ:cont_M}) we have that if the conduction process is present in different materials, the solution is made by taking Eq.(\ref{EQ:h_ast}) for each material and finally joining the solutiosn by requesting the continuity in both $j_M$ and $H^*$.

\subsubsection{Adiabatic conditions}
\label{IIIA2}

When the temperature is not externally controlled, we have to formulate the thermal problem by writing the heat diffusion equation. To this aim we need to write the continuity equations for the entropy. In stationary conditions Eq.(\ref{EQ:cont_s}) becomes $\nabla_x j_s = \sigma_s$ where the term at the left hand side is written by using $j_s=j_q/T$ and Eq.(\ref{EQ:jq_const}) rewritten as

\beq
j_q = \epsilon_{M} T j_{M} - \kappa \nabla_x T
\label{EQ:q_currjM}
\eeq

\noindent while the term at the right hand side is given by Eq.(\ref{EQ:sigma_s}). After a few passages, we obtain

\beq
\nabla_x j_q =  \mu_0\nabla_x H^* j_M + \frac{\mu_0\left( H^* \right)^2}{\tau_{M}}
\eeq

\noindent where the terms at the right hand side are due to the energy dissipation of the magnetization current and to the local damping, respectively. Both terms are quadratic in the force and the potential, therefore if we assume small currents and forces we are allowed to neglect them in a first approximation. In this case we obtain the condition $\nabla_x j_q =0$ which, in one dimension, corresponds to a constant heat flux traversing the material. Moreover we choose here to study the adiabatic condition corresponding to $j_q=0$ in which the two terms at the right hand side of Eq.(\ref{EQ:q_currjM}), the spin Peltier term $\epsilon_{M} T j_M$ and the heat conduction caused by the temperature profile $T(x)$, counterbalance each other, giving no net heat flow through the layer. The profile $T(x)$ will be stable if temperature of the thermal baths at the boundaries of the material are let free to adapt at the temperatures of the two ends. By using the adiabatic condition $j_q=0$ in Eq.(\ref{EQ:jq_const}) we immediately obtain 

\beq
\nabla_x T = \frac{1}{\hat{\epsilon}_{M}} \mu_0 \nabla_x H^*
\label{EQ:nablaT_ad}
\eeq

\noindent where $\hat{\epsilon}_{M}$ is the thermomagnetic power coefficient in adiabatic conditions

\beq
\frac{1}{\hat{\epsilon}_{M}} = \frac{1}{\epsilon_{M}} \frac{\kappa_M}{\kappa + \kappa_M}
\label{EQ:hat_epsilon}
\eeq

\noindent and $\kappa_M = \epsilon_{M}^2 \sigma_{M} T$. From Eq.(\ref{EQ:nablaT_ad}) we see that the temperature profile depends on the profile of the potential $H^*$. This last one is determined by inserting Eq.(\ref{EQ:nablaT_ad}) into Eq.(\ref{EQ:jM_const}). We have finally

\beq
j_M = \hat{\sigma}_{M} \mu_0 \nabla_x H^*
\label{EQ:}
\eeq

\noindent that has to be solved with the continuity equation (\ref{EQ:cont_M}) giving again the diffusion equation (\ref{EQ:diff_eq}) of the previous section. However now the diffusion length is the adiabatic value $\hat{l}_{M}=(\mu_0\hat{\sigma}_{M}\tau_{M})^{1/2}$ where 

\beq
\hat{\sigma}_{M} = \sigma_{M} \frac{\kappa}{\kappa+\kappa_M}
\eeq

\noindent is the conductivity for the magnetization current in adiabatic conditions.

\subsection{Spin Hall effect in non-magnetic metals}
\label{IIIB}
The spin Hall effect is due to the spin orbit interaction for conduction electrons. This effect is particularly relevant for noble metals with high atomic number. Because of the spin orbit interaction, a spin polarized electric current is deflected by an angle which is called the spin Hall angle $\theta_{SH}$. To include spin Hall effects into the theory of Section \ref{IIIA} one should first extend the equations for the thermo-magnetic effects to the presence of an additional electric current. This is straightforward and the formal result is reported in Appendix A. However to state the equation for the spin Hall effect, the equations must be further extended for two dimensional flow. The complete constitutive equations are characterized by six force variables, namely: the derivative along $x$ and $y$ of the three driving forces for magnetic, electric and heat currents. Here we simplify the problem by just disregarding the thermal effects. For our final aims this is a reasonable approximation, since the contribution arising from the  thermomagnetic coefficients of Pt is smaller than the other contributions involved in the full matrix of the thermo-magneto-electric effects\cite{Kikkawa-2013}. The general constitutive equations for the joint electric and magnetic transport are reported in Appendix B. Here we analyze in more detail the case of a non magnetic conductor with negligible Hall effect. We select the conditions in which the electric current $j_{e}$ is always along $y$, and the magnetization current $j_{M}$ along $x$. We have then the equations for the spin Hall and the inverse spin Hall effects from Eqs.(\ref{EQ:jeyb}) and (\ref{EQ:jmxb}). By converting to magnetic units one obtains

\bea
\label{EQ:jey}
j_{ey} & = & - \sigma_0 \nabla_{y} V_{e} + \sigma_0 \theta_{SH} \left(\frac{\mu_B}{e} \right) \mu_0 \nabla_{x} H^* \\
\label{EQ:jMx}
j_{Mx} & = &  \sigma_0 \theta_{SH} \left(\frac{\mu_B}{e} \right) \nabla_{y} V_e + \sigma_M \mu_0 \nabla_{x} H^*
\eea

\noindent where $\sigma_M = \sigma_0 ({\mu_B}/{e})^2$ is the conductivity for the magnetization current, $\sigma_0$ is the electric conductivity, $V_e$ is the electric potential, $e$ is the elementary charge and $\mu_B$ is the Bohr magneton. The equations contain the spin Hall effects in the non diagonal terms which couples different directions and different currents. It is worthwhile to notice that the effects are fully described by the spin Hall angle $\theta_{SH}$ which for metals is a definite negative quantity.

\subsubsection{Spin Hall effect}

In the spin Hall effect a magnetization current is generated in the parallel direction $x$ because of an electric current in the perpendicular one $y$. By eliminating $\nabla_{y} V_{e}$ by Eq.(\ref{EQ:jey}) and Eq.(\ref{EQ:jMx}) we find that the magnetization current is related to the electric current density by

\beq
j_{Mx} =-\theta_{SH}\left(\frac{\mu_B}{e}\right)j_{ey} +\sigma_{M}^{\prime}\mu_0 \nabla_{x} H^*
\label{EQ:jM_spinHall}
\eeq

\noindent where $\sigma_{M}^{\prime}=\sigma_M(1+\theta_{SH}^2)$. If the electric current density is uniform, the spin Hall effect corresponds to a magnetization current source $j_{MS}=-(\mu_B/e)\theta_{SH}j_{ey}$. The profile of the magnetization current $j_{Mx}$ which is actually traversing the layer also depends on the boundary conditions posed by the adjacent layers. Then, to find the profile $j_{Mx}(x)$, Eq.(\ref{EQ:jM_spinHall}) must be solved together with the continuity equation (\ref{EQ:cont_M}) giving a differential equation for the driving potential $H^{*}(x)$ which has the same from of Eq.(\ref{EQ:diff_eq}) but with $l_M= (\mu_0 \sigma_M' \tau_M)^{1/2}$. 

\subsubsection{Inverse spin Hall effect}

In the configuration corresponding to the inverse spin Hall effect one has a magnetization current in the parallel direction which generates an electric effect perpendicular to it. The electric equation in the $y$ direction is

\beq
j_{ey} = - \sigma_0^{\prime} \nabla_y V_{e} + \theta_{SH} \left(\frac{e}{\mu_B}\right) j_{Mx}
\label{EQ:je_invspinHall}
\eeq

\noindent where $\sigma_{0}^{\prime}=\sigma_0(1+\theta_{SH}^2)$. The magnetization current traversing the layer is not constant and it will be given by the solution of Eq.(\ref{EQ:jM_spinHall}) if the electric current $j_{ey}$ is constrained or by the solution of Eq.(\ref{EQ:jMx}) if the electric potential $\nabla_{y} V_{e}$ is constrained. In both cases the constitutive equation must be solved together with the continuity equation (\ref{EQ:cont_M}), giving again the differential equation (\ref{EQ:diff_eq}).

\section{Solutions of the magnetization current problem}

\subsection{Single active material}

\begin{figure}[htb]
\narrowtext 
\centering
\includegraphics[width=0.45\textwidth]{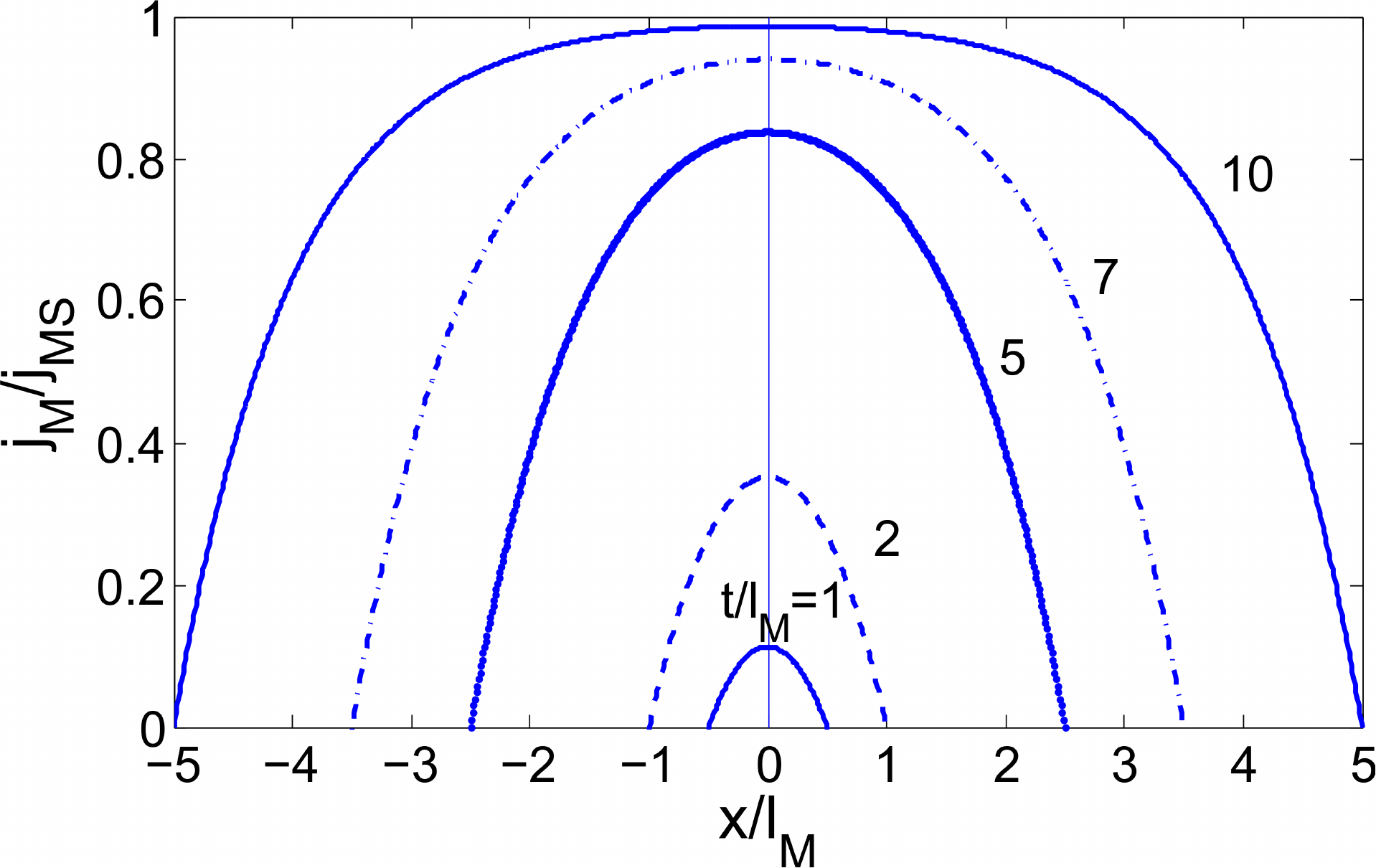}
\caption{Magnetization current profiles for a single active material. Curves are Eq.(\ref{jm_general}) with $d_1=-t/2$ and $d_2=t/2$,  boundary conditions fixed to zero ($j_M(-t/2)=j_M(t/2)=0$) and show different thicknesses $t/l_M$. The curves are normalized to $j_{MS}$.}
\label{FIG:general1}
\end{figure}

\begin{figure}[htb]
\narrowtext 
\centering
\includegraphics[width=0.45\textwidth]{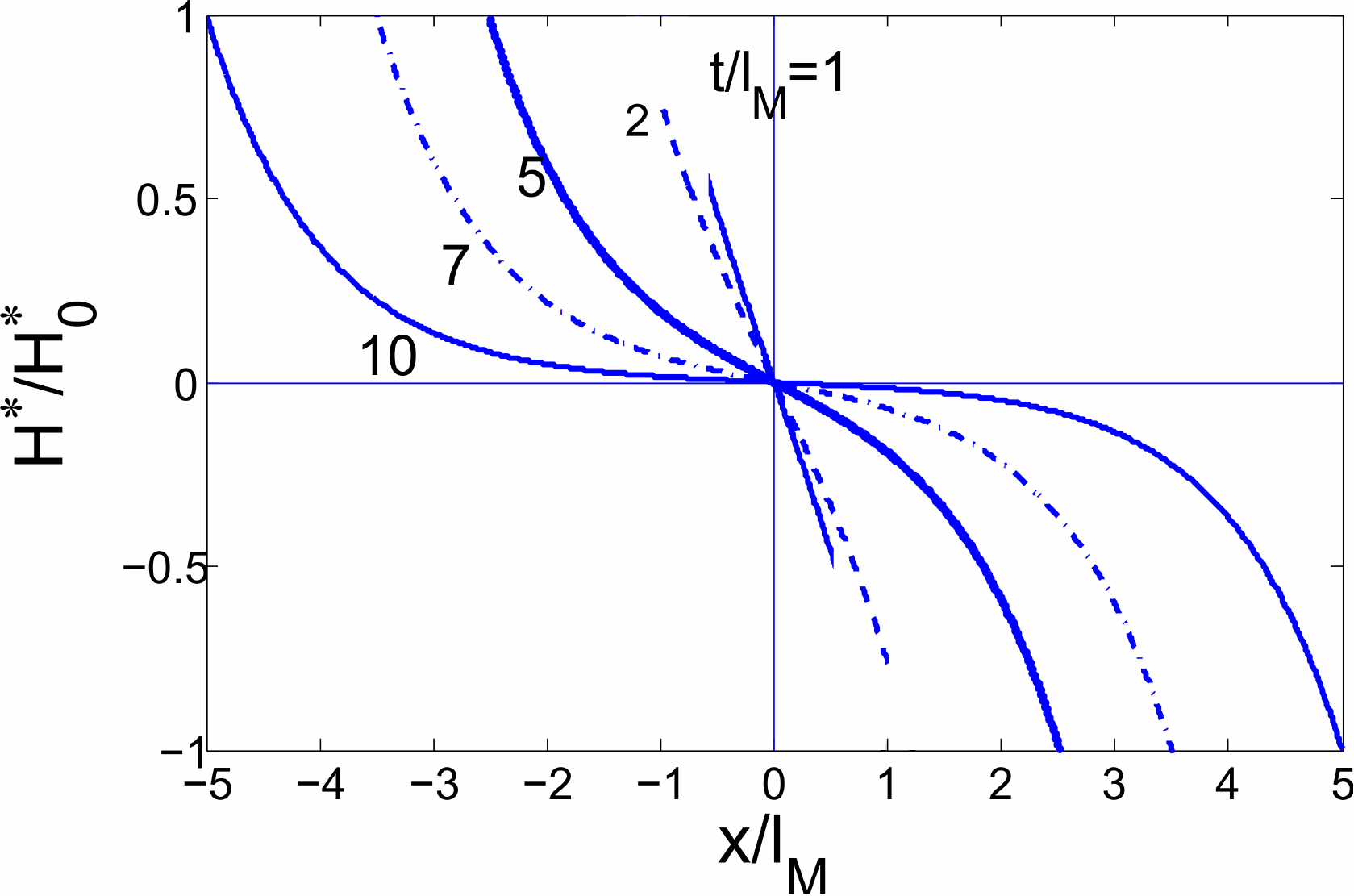}
\caption{Magnetization potential profile $H^*$ for a single active material. Curves are Eq.(\ref{H_general}) with $d_1=-t/2$ and $d_2=t/2$, boundary conditions fixed to zero ($j_M(-t/2)=j_M(t/2)=0$) and show different thicknesses $t/l_M$ (same as Fig.\ref{FIG:general1}). The curves are normalized to $H_0^{\ast}=j_{MS}/(l_M/\tau_M)$.} 
\label{FIG:general2}
\end{figure}

For an active material both the spin Seebeck effects and the spin Hall effect results in a magnetization current source and the profile of the magnetization current will be due to the boundary conditions. In presence of boundaries blocking the flow of the magnetization current, the magnetic moments accumulate giving rise to the potential $H^*$. The magnetization current close to a boundary is therefore absorbed by the materials itself as the potential $H^*$ is also the driving force for the non conservation of the magnetic moment (Eq.(\ref{EQ:dM_dt})). As it was shown in the previous Section, both spin Seebeck and spin Hall effects are characterized by constitutive equations that have the same functional form. Then we can work out the solution for the profile of the magnetization current independently of the specific effect and considering boundary conditions only. The specific solution will correspond to use as the current source $j_{MS}$ the expression derived from the spin Seebeck Eq.(\ref{EQ:jM_const}) or to the spin Hall Eq.(\ref{EQ:jM_spinHall}). We initially consider a single material with generic boundary conditions. The solution of the magnetization current problem with several layers will then be obtained by applying appropriate boundary conditions and joining the solutions for different layers. We take a material from $x=d_1$ to $x=d_2$ with a uniform source of magnetization current $j_{MS}$. Starting from the formal solution Eq.(\ref{EQ:h_ast}), we derive the magnetization current by Eq.(\ref{EQ:jM_isoT}) and we fix arbitrary values of the current at both boundaries, i.e. $j_M(d_1)$ and $j_M(d_2)$. The expression for the current is

\beq
\begin{split}
j_M(x) =& j_{MS} - (j_M(d_1)-j_{MS}) \frac{\sinh((x-d_2)/l_M)}{\sinh(t/l_M)} +\\
&+ (j_M(d_2)-j_{MS}) \frac{\sinh((x-d_1)/l_M)}{\sinh(t/l_M)}
\label{jm_general}
\end{split}
\eeq

\noindent and for the potential is

\beq
\begin{split}
H^*(x) =& - (j_M(d_1)-j_{MS}) \frac{1}{(l_M/\tau_M)}\frac{\cosh((x-d_2)/l_M)}{\sinh(t/l_M)} +\\
&+ (j_M(d_2)-j_{MS})\frac{1}{(l_M/\tau_M)}\frac{\cosh((x-d_1)/l_M)}{\sinh(t/l_M)},
\label{H_general}
\end{split}
\eeq

\noindent where $t=d_2-d_1$. Figs.\ref{FIG:general1} and \ref{FIG:general2} shows the profiles of the magnetization current and the effective field along the material for different thicknesses $t/l_M$. The spin accumulation close to the boundaries generates, as a reaction, an effective field which counteracts the effect considered (e.g. the spin Seebeck effect) in order to let the current to go to zero at the interface. 

\subsection{Injection of a magnetization current}

We consider the injection of a magnetization current from an active material which is acting as current generator, or current injector, into a passive material which is acting as a conductor. It is known that the quality of the interface plays an important role in the injection of the spin currents \cite{Qiu-2015}. In Ref.\cite{Qiu-2015} the condition of the Pt/YIG interface was intentionally modified by creating a thin amorphous YIG layer varying from 1 to 14 nm and it was shown that the spin Seebeck effect is depressed as the thickness of the amorphous layer increases. The maximum value is obtained with a fully crystalline interface and the typical decay length of the effect with thickness is $2.3$ nm. In the present theory this kind of interlayer interface can be taken into account by introducing a third effective layer, with degraded properties, between the two. In the present paper we consider ideal interfaces between injector and conductor which is appropriate for spin Seebeck experiments characterized by crystalline interfaces. To analyze the injection of a magnetization current, we simplify the notation by dropping the $M$ subscript and employing subscripts describing the role of the material: (1) for the injector and (2) for the conductor. The magnetization current source is that of the active material (1) and is denoted $j_{MS}$. The connection between the two media is set at $x=0$. The boundary conditions for the magnetization current is $j_1(0)=j_2(0)=j_0$ and the boundary condition for the potential is $H^*_1(0)=H^*_2(0)=H^*_0$. Appendix C reports the formal solutions in the case in which each layer has finite width. These solutions will be employed in the comparison with real experiments performed in bilayers. Here we discuss how the efficiency of the injections is determined by intrinsic parameters. To this aim we take the solutions of Appendix C in the limit of semi infinite width and we obtain

\beq
j_1(x)=j_{MS}-(j_{MS}-j_0) \exp(x/l_1)
\eeq

\noindent and 

\beq
j_2(x)=j_{0} \exp(-x/l_2)
\eeq

\noindent for the currents and 

\beq
H^{\ast}_1(x)=\frac{j_0-j_{MS}}{(l_1/\tau_1)}\exp(x/l_1)
\eeq

\noindent and 

\beq
H^{\ast}_2(x)= -\frac{j_0}{(l_2/\tau_2)}\exp(-x/l_2).
\eeq

\noindent By setting the boundary condition at the interface between the two media $H^{\ast}_1(0)=H^{\ast}_2(0)$ we find the value of the current at the interface

\beq
j_0 = \frac{j_{MS}}{1+r_{12}}
\eeq

\noindent where $r_{12}=(l_1/\tau_1)/(l_2/\tau_2)$. If $r_{12}\ll1$ the current is efficiently injected, while if $r_{12}\gg1$ the magnetization current is not transmitted into the conductor. In terms of intrinsic parameters we have

\beq
r_{12} = \sqrt{\frac{\sigma_1}{\sigma_2}\frac{\tau_2}{\tau_1}}.
\eeq

\noindent So a junction with an efficient injection from (1) to (2) should have a conductor (2) with a magnetization conductivity much larger than the injector $\sigma_2 \gg \sigma_1$ and a time constant much smaller $\tau_2 \ll \tau_1$. 

\section{Spin Seebeck and Spin Peltier effects}

In this Section we apply the theory previously developed to the spin Seebeck and spin Peltier effects.

\subsection{Spin Seebeck effect}

\begin{figure}[htb]
\centering
\includegraphics[width=0.45\textwidth]{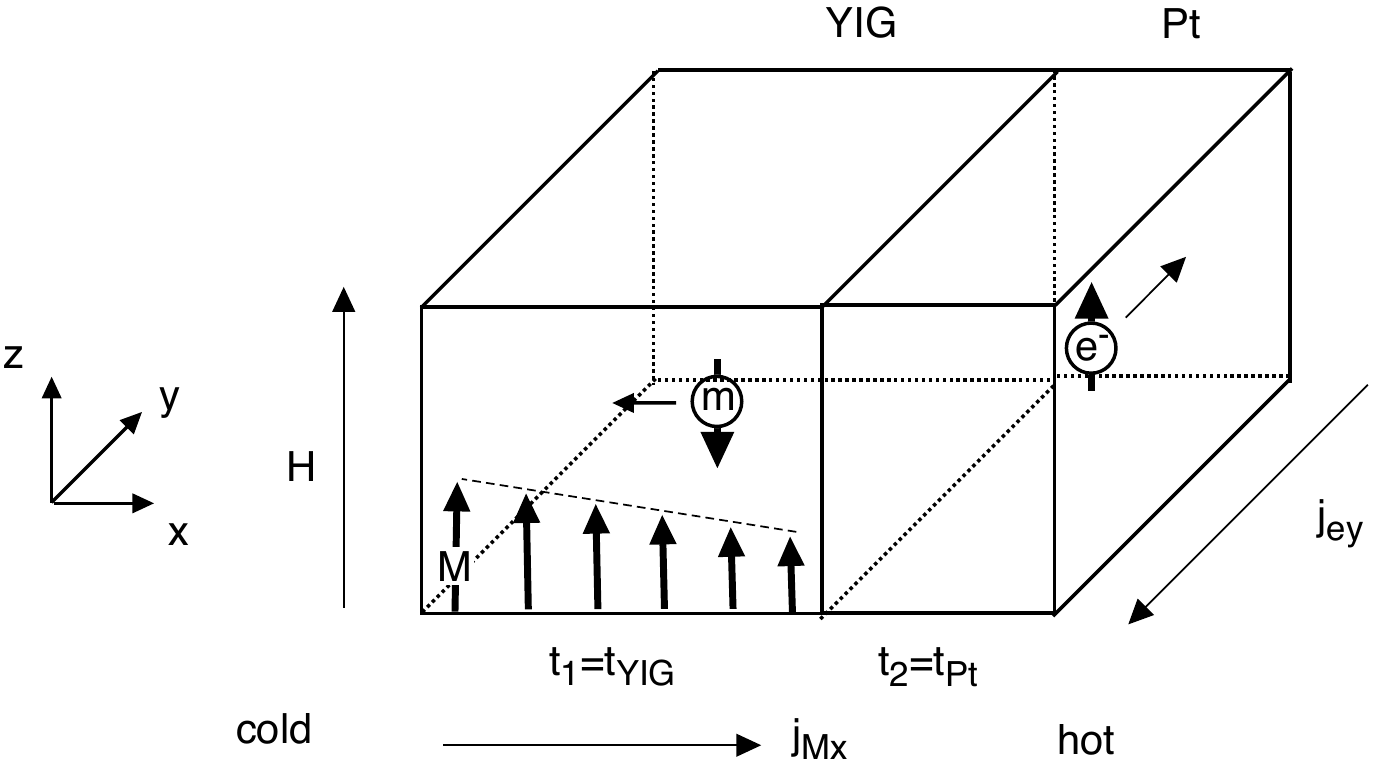}
\caption{Geometry of the longitudinal spin Seebeck effect. 
} \label{FIG:spinSeebeck}
\end{figure}

The spin Seebeck effect consists in a magnetization current generated by a temperature gradient across a ferromagnetic material. We study the longitudinal spin Seebeck effect (LSSE) where the magnetization current and the temperature gradient are along the same direction. We consider experiments in which the active layer is YIG, the injector, labeled as (1) and the sensor layer is Pt, the conductor, labeled as (2). The geometry of the experiment is schematically shown in Fig.\ref{FIG:spinSeebeck}. The YIG injector has thickness $t_1=t_{YIG}$ while the Pt conductor has thickness $t_2=t_{Pt}$. The interface is set at $x=0$. 

The temperature gradient is applied along $x$, the magnetic field is along $z$, the electric effects (ISHE voltage) are  measured along $y$. We consider a constant temperature gradient $\nabla_x T$, therefore the magnetization current source of YIG is $j_{MS} = - \sigma_{YIG} \epsilon_{YIG} \, \nabla_x T$ given by the equations of Section \ref{IIIA}. The solutions of the magnetization current problem are Eqs.(\ref{EQ:AC_j1}) and (\ref{EQ:AC_j2}) reported in Appendix C and the magnetization current at the interface is given by Eq.(\ref{EQ:AC_j0}) in which $l_1=l_{YIG}$, $\tau_1=\tau_{YIG}$ and $l_2=l_{Pt}$, $\tau_2=\tau_{Pt}$. As the thickness of the Pt layer is generally of the same order of the spin diffusion length ($t_{Pt} \sim l_{Pt} \sim 10$ nm), we can approximate Eq.(\ref{EQ:AC_j2}) for the case of $t_2 \sim l_2$ and find that the profile of the magnetization current is, at a good approximation, a linear decay from $j_0$ at the interface $x=0$ to zero at the border $x=t_2$. The average magnetization current in the Pt layer is therefore $\langle j_{Mx}\rangle_x = j_0/2$ where $j_0$ is the magnetization current injected at the interface. If the experiments are performed by measuring the ISHE voltage, by taking Eq.(\ref{EQ:je_invspinHall}) with $j_{ey}=0$, we obtain the relation between the magnetizations current along $x$ and the electric potential along $y$. We assume the relation to be valid for the average values along $x$ over the thickness $t_2$. The average potential is then

\beq
\langle\nabla_y V_e\rangle_x = \frac{\theta_{SH}}{\sigma_{e}} \left(\frac{e}{\mu_B}\right) \langle j_{Mx}\rangle_x. 
\eeq

\noindent where $\sigma_{e}$, corresponding to $\sigma_0^{\prime}$ in Eq.(\ref{EQ:je_invspinHall}), is the electric conductivity of Pt. The current injected at the interface $j_0$ can therefore be estimated by the gradient of the ISHE voltage $\nabla_y V_{ISHE}= \langle \nabla_y V_{e}\rangle_x$, 

\beq
j_0 = 2 \frac{\sigma_e}{\theta_{SH}}\left(\frac{\mu_B}{e}\right)\nabla_y V_{ISHE}. 
\label{EQ:j0_LSSE} 
\eeq

\noindent In experiments, the spin Seebeck coefficient is determined as $S_{LSSE} = \nabla_yV_{ISHE}/\nabla_xT$. The magnetization current at the interface can be calculated by Eq.(\ref{EQ:j0_LSSE}) where the spin Hall angle is evaluated as $\theta_{SH} =-0.1$ from Ref.\cite{Wang-2014}. In turn, the relation between the spin Seebeck current $j_{MS}$ and $j_0$ at the interface, given by Eq.(\ref{EQ:AC_j0}), will depend on the intrinsic parameters of both layers and their thickness. Once the current $j_{MS}$ is calculated, one can estimate the spin Seebeck coefficient as

\beq
\epsilon_{YIG} = \frac{1}{\sigma_{YIG}}\left( \frac{j_{MS}}{-\nabla T}\right).
\eeq

\noindent  In Pt the magnetization diffusion length is known to be $l_{Pt}=7.3$ nm\cite{Wang-2014}. The spin conductivity can be estimated by assuming that in a normal metal the scattering acts independently of the spin \cite{Valet-1993}. Then, by converting the electrical conductivity of Pt $\sigma_{e} = 6.4 \cdot 10^6 \, \Omega^{-1}$m$^{-1}$, into the conductivity for the magnetization current, we obtain $\mu_0\sigma_{Pt}=2.6 \cdot 10^{-8}$ m$^{2}$s$^{-1}$. The time constant is finally calculated and results $\tau_{Pt} = l_{Pt}^2/(\mu_0\sigma_{Pt}) \simeq 2 \cdot 10^{-9}$ s. 

In YIG the estimations of the magnetization diffusion length present in literature, range from micron to millimeter \cite{Uchida-2012, Giles-2015, Cornelissen-2015} for the transverse experiment (in which current and magnetization are parallel) to much lower value (i.e. $< 1 \mu$m) \cite{Ramos-2015} for the longitudinal effect (in which current and magnetization are perpendicular). From Ref.\cite{Uchida-2014} the LSSE coefficient measured on 1 mm of YIG, $S_{LSSE} \simeq 4\cdot10^{-7}$ VK$^{-1}$, results to be larger than the one measured on a 4 $\mu$m sample\cite{Sola-2015} $S_{LSSE} \simeq 2.8\cdot10^{-7}$ VK$^{-1}$, but of the same order of magnitude. Therefore we can guess that $l_{YIG}$ is of the same order of magnitude of the thinner sample (4 $\mu$m) in order to allow for an efficient injection in both cases. In a more recent study, the dependence of the spin Seebeck effect on the thickness of YIG was investigated \cite{Kehlberger-2015}. It has been reported that the typical diffusion length is below $l_{YIG} = 1.5 \, \mu$m. We set in the following $l_{YIG} = 1 \, \mu$m. For the evaluation of the absolute thermomagnetic power coefficient $\epsilon_{YIG}$ we use the result of Ref.\cite{Sola-2015} where the thermal conditions were properly taken into account. These experiments were performed by using a YIG layer of 4 $\mu$m and a Pt layer of 10 nm. 

By using the LSSE coefficient estimated at the saturation magnetization of YIG we obtain $j_0/(-\nabla_x T) \simeq  2 \cdot10^{-3}$ As$^{-1}$K$^{-1}$m \cite{Sola-2015}. The only missing intrinsic parameter is the magnetization conductivity of the YIG, $\sigma_{YIG}$. To have an order of magnitude we suppose a reasonable injection from YIG into Pt (i.e 50\%, with $j_0= 0.5 \, j_{MS}$). Then we set $r_{12}=1$, i.e. $l_1/\tau_1 = l_2/\tau_2$. By using the resulting value for the magnetization conductivity of YIG $\mu_0\sigma_{YIG} \sim 4 \cdot 10^{-7}$ m$^{2}$s$^{-1}$, we finally obtain an order of magnitude for the absolute thermomagnetic power coefficient as $\epsilon_{YIG} \sim 10^{-2}$ TK$^{-1}$. In analogy with the  thermoelectric effects where the absolute thermoelectric power coefficient is compared to the classical value $\epsilon_e = -k_B/e \simeq - 86 \cdot 10^{-6}$ VK$^{-1}$, the value found here can be compared with the ratio $k_B/\mu_B \simeq 1.49 $ TK$^{-1}$ \cite{Rezende-2014}. Furthermore, as the experiments show that $\nabla_y V_{ISHE}$ and therefore $j_{MS}$, changes sign when the magnetization of the YIG layer is inverted, this means that $\epsilon_{YIG}$ changes sign when inverting the magnetization $M$. The value reported before corresponds to the absolute value when the magnetization of YIG is at saturation.

\subsection{Spin Peltier effect}

\begin{figure}[htb]
\centering
\includegraphics[width=0.45\textwidth]{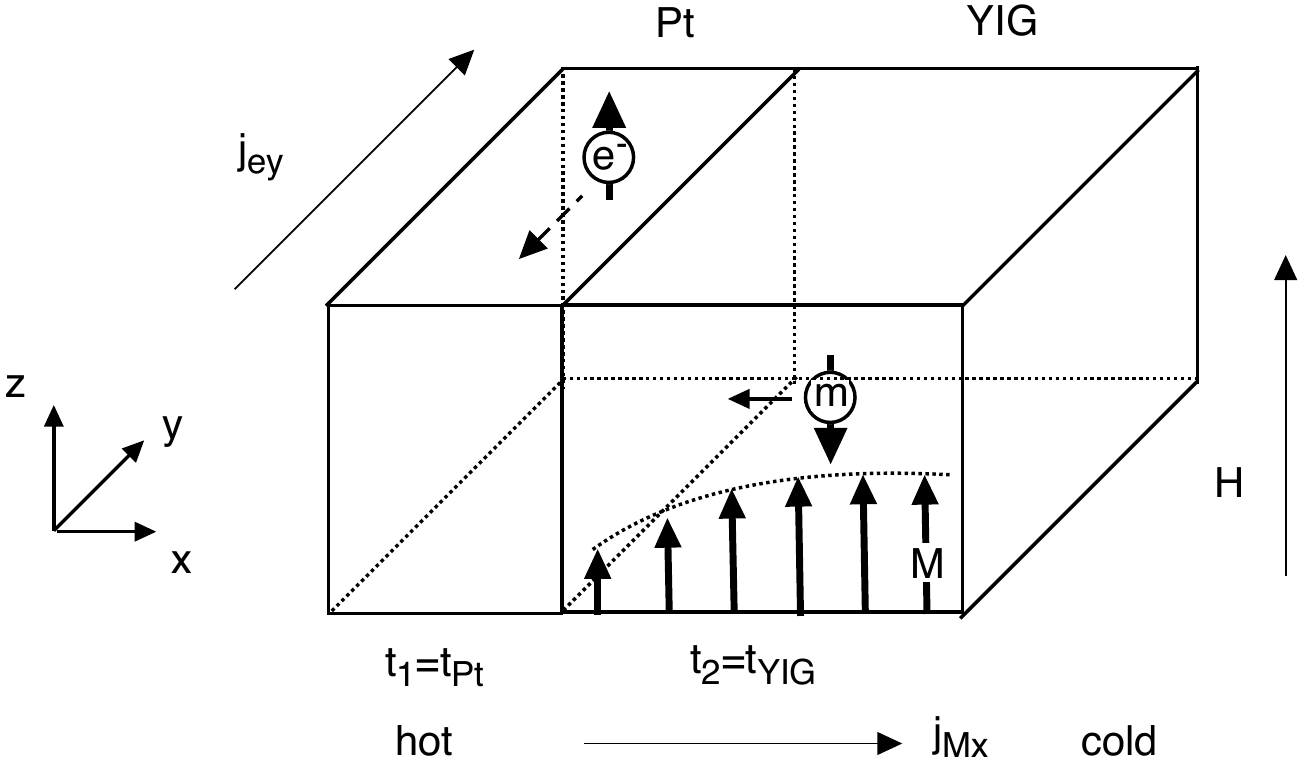}
\caption{Geometry of the spin Peltier effect. 
} \label{FIG:spinHallorigin}
\end{figure}

In the spin Peltier experiments a magnetization current is generated by the spin Hall effect in a Pt layer, labeled as (1) and is injected into a YIG layer, labeled as (2). The injection of the magnetization current into the YIG, generates thermal effects. The geometry of the experiment is schematically shown in Fig.\ref{FIG:spinHallorigin}. 

\begin{figure}[htb]
\narrowtext 
\centering
\includegraphics[width=0.45\textwidth]{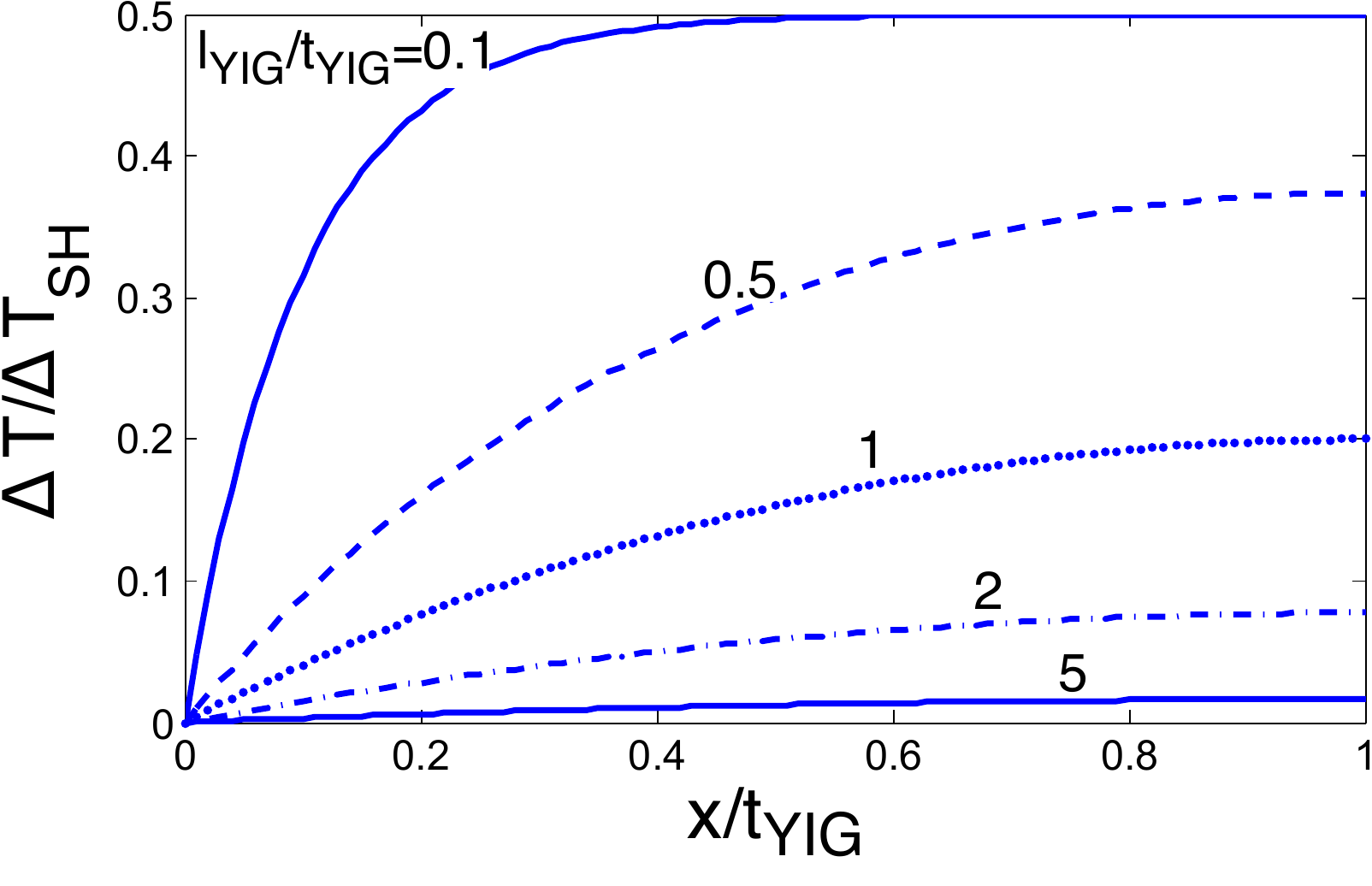}
\caption{Temperature profile of YIG for the spin Peltier effect. Curves are $\Delta T = T(x)-T_(0)$ from Eq.(\ref{EQ:deltaT}) normalized to $\Delta T_{SH}=\mu_0H_{SH}^{\ast}/\hat{\varepsilon}_{YIG}$ and $H_{SH}^{\ast} = j_{MS}/(l_{YIG}/\tau_{YIG})$. The parameters are $r_{12}=1,\, l_{Pt}/t_{Pt}=0.1$.} \label{FIG:t}
\end{figure}

The interface is set at $x=0$, the electric current is along $y$, the magnetization current is along $x$ and the magnetic field is along $z$. The magnetization current source is now $j_{MS} = -\theta_{SH}\left({\mu_B}/{e}\right)j_{ey}$ given by the spin Hall effect in Pt discussed in Section \ref{IIIB}. When the magnetization current diffuses inside YIG, it also generates a heat current because of the spin Peltier effect described in Section \ref{IIIA2}. The solution of the magnetization conduction problem is mathematically identical to the spin Seebeck one, but with the role of YIG and Pt inverted. For this reason we have employed label (1) for the injector, which is now Pt, and label (2) for the conductor which is now YIG. The solutions of the magnetization current problem are again Eqs.(\ref{EQ:AC_j1}) and (\ref{EQ:AC_j2}) reported in Appendix C and the magnetization current at the interface is given by Eq.(\ref{EQ:AC_j0}). With respect to the previous spin Seebeck case, the diffusion length of YIG is the adiabatic value $\hat{l}_{YIG}=(\mu_0\hat{\sigma}_{YIG}\tau_{YIG})^{1/2}$. In the spin Peltier experiment the temperature profile in YIG is given by the integration of Eq.(\ref{EQ:nablaT_ad})

\begin{equation}
T(x)-T(0)= \frac{1}{\hat{\epsilon}_{YIG}} \mu_0 \left(H_2^{\ast}(x)-H_2^{\ast}(0)\right)
\label{EQ:deltaT}
\end{equation}

\noindent where $H_2^{\ast}(x)$ is given by Eq.(\ref{EQ:AC_h2}). The result is shown in Fig.\ref{FIG:t}.

By looking at the magnetization current profile (Fig.\ref{FIG:junction1}), we see, as in the spin Seebeck experiment, that in order to have a good efficiency, the thickness of each layer should be larger than its diffusion length ($t_1 > l_1$ and $t_2 > l_2$) to permit to the magnetization current to develop. Moreover the efficiency of the injection is regulated by the ratio of intrinsic parameters $r_{12} = (l_1/\tau_1)/(l_2/\tau_2)$, where $(1)$ is the injector Pt and $(2)$ is the conductor YIG. Again the magnetization current at the interface is large if the ratio $r_{12}$ is small. However it should be noticed that given the two materials in the junction (i.e. Pt,YIG) we have that $r_{Pt \rightarrow YIG} = 1/ r_{YIG \rightarrow Pt}$. So, the value $r_{Pt \rightarrow YIG} = r_{YIG \rightarrow Pt} \simeq 1$ is the value which permits relatively efficient injection both from Pt into YIG and from YIG into Pt.

Finally from the temperature profile Fig.\ref{FIG:t} obtained in adiabatic conditions we can reach information about the coefficient of the absolute thermomagnetic power in adiabatic conditions $\hat{\varepsilon}_{YIG}$. The profile $T(x)$ is normalized to the temperature $\Delta T_{SH}$ which gives the typical scale of the effect  

\beq
\Delta T_{SH}=\frac{1}{\hat{\varepsilon}_{YIG}} \mu_0H_{SH}^{\ast}
\eeq

\noindent where $H_{SH}^{\ast} = j_{MS}/(l_{YIG}/\tau_{YIG})$. From the literature the thermal conductivity of YIG is $\kappa = 6$ W K$^{-1}$m$^{-1}$. From Section V A, $\varepsilon_{YIG} \simeq 10^{-2}$ TK$^{-1}$ and the parameter $\kappa_{YIG}\simeq 10^{-2}$ W K$^{-1}$m$^{-1}\,$\cite{Basso-2015}. Moreover the potential $H_{SH}^{\ast}$ is related to the spin Hall current $j_{MS}=-(\mu_B/e)\theta_{SH}j_{ey}$ injected from Pt. Using the values from \cite{Basso-2015} $l_{YIG}/\tau_{YIG}=3$ ms$^{-1}$ and $\theta_{SH}=-0.1$ we are able to give an order of magnitude estimate of the temperature change, obtaining $\Delta T_{SH}/j_{ey} = 4 \cdot 10^{-13}$ K A$^{-1}$m$^2$. 

Experimental values are taken from Ref.\cite{Flipse-2014}, where in correspondence to an electric current density of $3 \cdot 10^{10}$ A m$^{-2}$ in Pt, the temperature difference measured by a thermocouple in YIG was $2.5 \cdot 10^{-4}$ K, considering that the Joule heating of the electric current in Pt was already subtracted. The parameter $\Delta T_{SH}$ results $1.2 \cdot 10^{-2}$ K which is of the correct order of magnitude. Consequently by using $t_1=t_{Pt} = 5$ nm and $t_2=t_{YIG} = 0.2\, \mu$m in Eqs.(\ref{EQ:AC_h2}) and (\ref{EQ:deltaT}), we find an adiabatic temperature change of $T(t_{YIG})-T(0) \simeq 2.5 \cdot 10^{-4}$ K with $l_{YIG}=0.4\, \mu$m. This value refines the upper limit of $1\, \mu$m which was found in Section V A, however the phenomenology of the spin Peltier effect in YIG seems coherent with the absolute thermomagnetic power coefficient derived previously.

\section{Conclusions}

In this paper the problem of magnetization and heat currents is investigated through a non equilibrium thermodynamics approach. Based on the constitutive equations of a ferromagnetic insulator and a spin Hall active material we are able to solve the problem of the profiles of the magnetization current and of the potential in the geometry of the longitudinal spin Seebeck and of the spin Peltier effects.
By focusing on the specific geometry with one YIG layer and one Pt layer, we obtain the optimal conditions for generating large magnetization currents into Pt in the case of the spin Seebeck effect and for generating large heat current in YIG in the case of spin Peltier effects. In both cases we find that efficient injection is obtained when the thickness of the injecting layer is larger than the diffusion length $l_M$. The theory predictions are compared with experiments and this permits to determine the values of the thermomagnetic coefficients of YIG: the magnetization diffusion length $l_M \sim 0.4 \, \mu$m and the absolute thermomagnetic power coefficient $\epsilon_M \sim 10^{-2}$ TK$^{-1}$.

\section*{Acknowledgments}
This work has been carried out within the Joint Research Project EXL04 (SpinCal), funded by the European Metrology Research Programme. The EMRP is jointly funded by the EMRP participating countries within EURAMET and the European Union.

\appendix 

\section{Constitutive equations of the thermo-magneto-electric effects}

The equations for the thermo-magneto-electric effects relates the current densities of the electric charge $\boldsymbol{j}_e$, the magnetic moment $\boldsymbol{j}_M$, and the heat $\boldsymbol{j}_q$, with the gradients of the electric potential $V_e$, the magnetization potential $H^*$, and the temperature $T$. In one dimension ($\nabla_x = \partial/\partial x$) the equations are

\begin{align}
j_e & = -\sigma_e \, \nabla_x V_e + \eta \, \mu_0 \nabla_x H^* - \sigma_e \epsilon_e  \, \nabla_x T \label{EQ:jconstit1}\\
j_M & = - \eta \, \nabla_x V_e + \sigma_M \, \mu_0 \nabla_x H^* - \sigma_M \epsilon_M  \, \nabla_x T \label{EQ:jconstit2}\\
j_q & = - \epsilon_e \sigma_e T \, \nabla_x V_e + \epsilon_M \sigma_M T \mu_0 \nabla_x H^* - \kappa_{iso}  \nabla_x T. \label{EQ:jconstit3}
\end{align}

\noindent where $\sigma_e$ is the electrical conductivity, $\epsilon_e$ is the absolute thermoelectric power coefficient, $\eta$ represents the magneto-electric conductivity, $\sigma_M$ is the magnetic conductivity, $\epsilon_M$ is the absolute thermomagnetic power coefficient and $\kappa_{iso}$ is the thermal conductivity with $\nabla_x V_e=0$ and $\nabla_x H^*=0$. By defining the heat current as $j_{q} = T j_{s}$ we obtain from Eqs.(\ref{EQ:cont_s}) and (\ref{EQ:sigma_s})

\beq
T\frac{\partial s}{\partial t} + \nabla_x j_q =  \mu_0 \nabla_x H^* j_M + \frac{\mu_0\left( H^* \right)^2}{\tau_M}- \nabla_x V_e j_e.
\label{EQ:heat}
\eeq

\noindent By solving the previous equation together with the constitutive equation (\ref{EQ:jconstit3}), one can obtain the generalized heat diffusion equations.

\section{Magnetization current carried by electrons}

We consider the specific case of metals in which the magnetic and electric current are due to the same type of carriers (electrons or holes) with different spin. The theory can be equivalently formulated in terms of magnetic moment (up or down). One subdivides the particle current $j_n = j_{n+} + j_{n-}$ into the sum of moment up $j_{n+}$ and moment down $j_{n-}$. The electric current is $j_e = q j_n$ where $q$ is the charge of the carrier, while the magnetization current is $ \mu_B (j_{n+} - j_{n-})$, where $\mu_B$ is the Bohr magneton. As it is  somehow customary to define a magnetization current $j_m$ measured in the same units of the electric currents, we have then $j_m = (e/q) (j_{e+} - j_{e-})$ where $e$ is the elementary charge. Electrons, moving in the opposite direction of the charge current, with a magnetic moment up will give a negative $j_m$, while holes with moment up, will give a positive $j_m$. One is allowed to assume different conductivities among the two sub-bands as a function of the gradients of the potentials $\nabla V_{e\pm}$ relative to each sub-band. The equations are

\bea
j_{e+} &=& -\sigma_{+} \nabla V_{e+}-\sigma_{mix} \nabla V_{e-}\\
j_{e-} &=& -\sigma_{mix} \nabla V_{e+}-\sigma_{-} \nabla V_{e-}
\eea

\noindent where one has to introduce both the individual channel conductivities $\sigma_{+}$ and $\sigma_{-}$ and the spin mixing conductivity $\sigma_{mix}$. One obtains

\beq
\label{EQ:system}
\left( 
\begin{array}{c} 
j_{e} \\
j_{m} \\
\end{array}
\right)
= - \sigma_{0} \left( 
\begin{array}{cc}
1+\alpha  & \beta \\
\beta  & 1-\alpha \\
\end{array}
\right)
\left( 
\begin{array}{c}
\nabla V_{e}\\
\nabla V_{m}\\
\end{array}
\right)
\eeq

\noindent with $V_e = V_{e+}+V_{e-}$ and $V_m = (e/q)(V_{e+}-V_{e-})$ where $\sigma_{0} =(\sigma_{+}+\sigma_{-})/2$ is the electric conductivity, $\alpha = \sigma_{mix}/\sigma_{0}$ is the spin mixing coefficient ($\alpha \le 1$) and $\beta = (\sigma_{+}-\sigma_{-})/(2\sigma_0)$ represents the spin unbalance of the conductivities. $V_m$ is a potential for the current $j_m$ with the same units of $V_e$. The electric conductivity is $\sigma_{e} = \sigma_{0} (1+\alpha)$ and the conductivity for the magnetization current $j_m$ is $\sigma_{m} = \sigma_{0} (1-\alpha)$. It is often the case that the spin mixing conductivity is very small (i.e.  $\alpha=0$ into Eq.(\ref{EQ:system})) because the spin flip events are much more rare than the normal scattering conserving the spin, so $\sigma_m = \sigma_e$. This leads to the Mott's two current model. In that case the spin unbalance coefficient $\beta$ is a number between 1 and -1. 


The previous equations form also the basis to describe the Hall and the spin Hall effects. We need to extend the equations for the magneto-electric effects to two dimensions. We consider the case in which the spin mixing conductivity is zero and $\sigma_e = \sigma_m=\sigma_0$. The equations read

\beq
\left( 
\begin{array}{c} 
j_{ex} \\ 
j_{ey} \\ 
j_{mx} \\
j_{my} 
\end{array}
\right)
= - \sigma_{0} \left( 
\begin{array}{cccc}
1  & -\theta_{H}      & \beta  &  -\theta_{SH} \\
\theta_{H}  & 1  & \theta_{SH} &  \beta \\
 \beta &  -\theta_{SH} & 1  & -\theta_{H} \\
\theta_{SH}  &  \beta & \theta_{H} & 1\\
 \end{array}
\right)
\left( 
\begin{array}{c}
\nabla_{x} V_{e}\\
\nabla_{y} V_{e}\\
\nabla_{x} V_{m}\\
\nabla_{y} V_{m}\\
\end{array}
\right)
\eeq

\noindent where $\theta_{H}$ is the Hall angle and $\theta_{SH}$ is the spin Hall angle. It is important to notice that the Hall angle depends on the magnetic field while the spin Hall angle is a constant that is determined by the spin orbit interaction for conduction electrons. 

We analyze in more detail a non magnetic conductor with $\beta=0$ for which the Hall angle is negligible $\theta_{H}=0$. Furthermore we select conditions in which the electric current is always along $y$ and the magnetic current along $x$. We have finally the equations for the spin Hall and the inverse spin Hall effects

\bea
\label{EQ:jeyb}
j_{ey}/\sigma_0 & = & - \nabla_{y} V_{e} - \theta_{SH} \nabla_{x} V_m \\
\label{EQ:jmxb}
j_{mx}/\sigma_0 & = &  \theta_{SH} \nabla_{y} V_e - \nabla_{x} V_m.
\eea

\noindent To convert to magnetic units of Section \ref{IIIB} one simply uses

\beq
\nabla V_{m} = - \left(\frac{\mu_B}{e} \right) \mu_0 \nabla H^* 
\eeq

\noindent and

\beq
j_m = \left(\frac{e}{\mu_B} \right) j_M. 
\eeq

\section{One junction}
Let us consider a bilayer of two materials: the injector (1) from $x=-t_1$ to $x=0$ which contains a magnetization current source $j_{MS}$ and the conductor (2) from $x=0$ to $x=t_2$. The connection between the two media is put at $x=0$ and the boundary conditions on the magnetization current are: $j_1(-t_1)=0$, $j_2(t_2)=0$ and $j_1(0)=j_2(0)=j_0$. The solutions for the magnetization currents, where only the injector (1) is an active material, are

\begin{multline}
j_1(x)=j_{MS}+j_{MS}\frac{\sinh(x/l_1)}{\sinh(t_1/l_1)}+\\+(j_0-j_{MS})[\sinh(x/l_1)\coth(t_1/l_1)+\cosh(x/l_1)]
\label{EQ:AC_j1}
\end{multline}

\noindent and

\begin{equation}
j_2(x)= - j_0 \, [\sinh(x/l_2)\coth(t_2/l_2)-\cosh(x/l_2)]
\label{EQ:AC_j2}
\end{equation}

\noindent and for the potentials

\begin{multline}
H^{\ast}_1(x)=\frac{j_{MS}}{(l_1/\tau_1)}\frac{\cosh(x/l_1)}{\sinh(t_1/l_1)}+\\+\frac{j_0-j_{MS}}{(l_1/\tau_1)}[\cosh(x/l_1)\coth(t_1/l_1)+\sinh(x/l_1)]
\label{EQ:AC_h1}
\end{multline}

\noindent and

\begin{equation}
H^{\ast}_2(x)= - \frac{j_0}{(l_2/\tau_2)}[\cosh(x/l_2)\coth(t_2/l_2)-\sinh(x/l_2)].
\label{EQ:AC_h2}
\end{equation}

\begin{figure}[htb]
\narrowtext 
\centering
\includegraphics[width=0.5\textwidth]{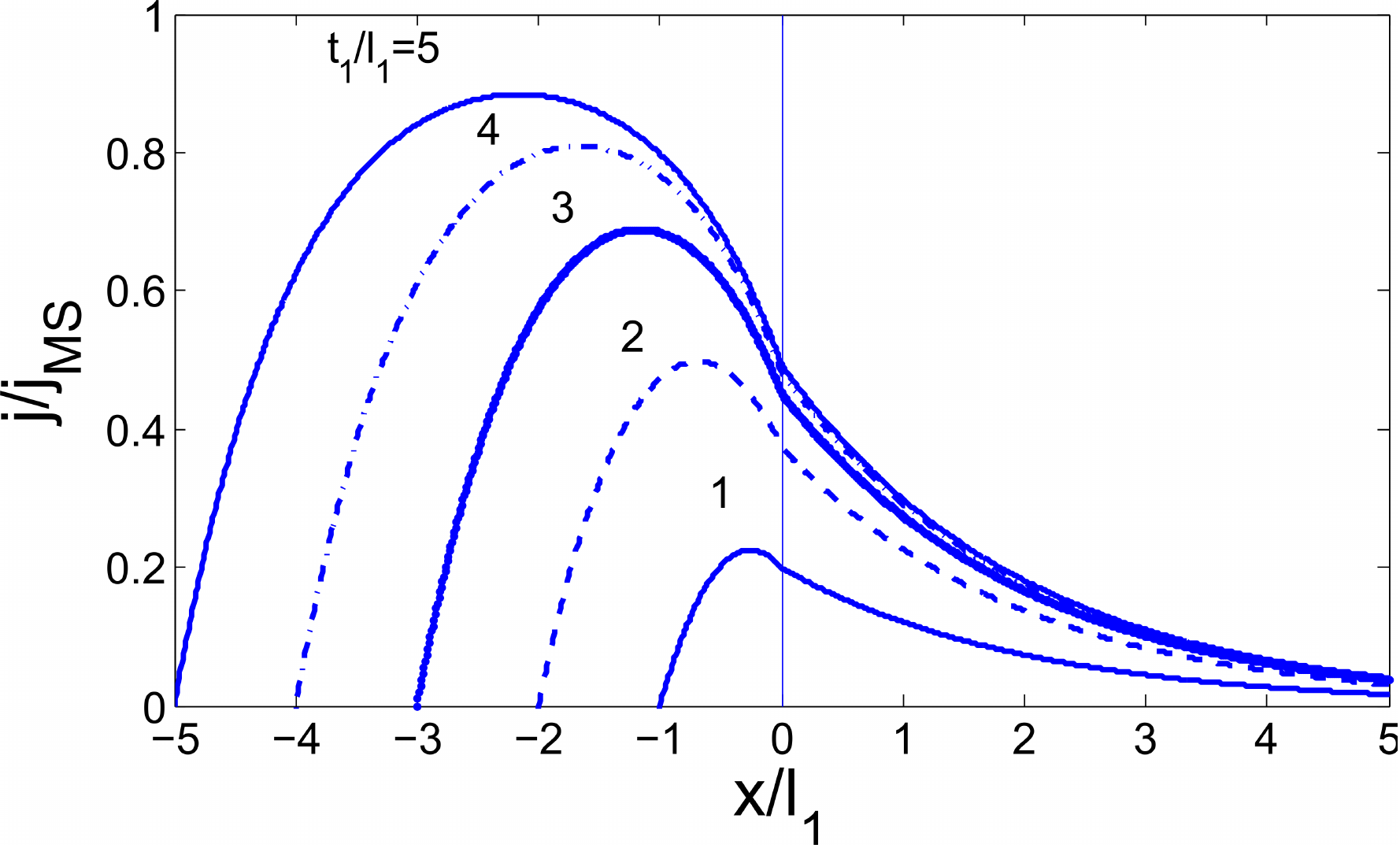}
\caption{Magnetization current profiles for a bilayer showing the passage (injection) of a magnetization current generated in medium (1), of finite thickness $t_1/l_1$, to the semi infinite conductor medium (2). Curves are from Eqs.(\ref{EQ:AC_j1}), (\ref{EQ:AC_j2}). The parameters are $r_{12}=1,\, l_2/l_1=2$. The curves show the effect of different thicknesses $t_1/l_1$ of layer (1) on the injected current.} 
\label{FIG:junction1}
\end{figure}

\begin{figure}[htb]
\narrowtext 
\centering
\includegraphics[width=0.5\textwidth]{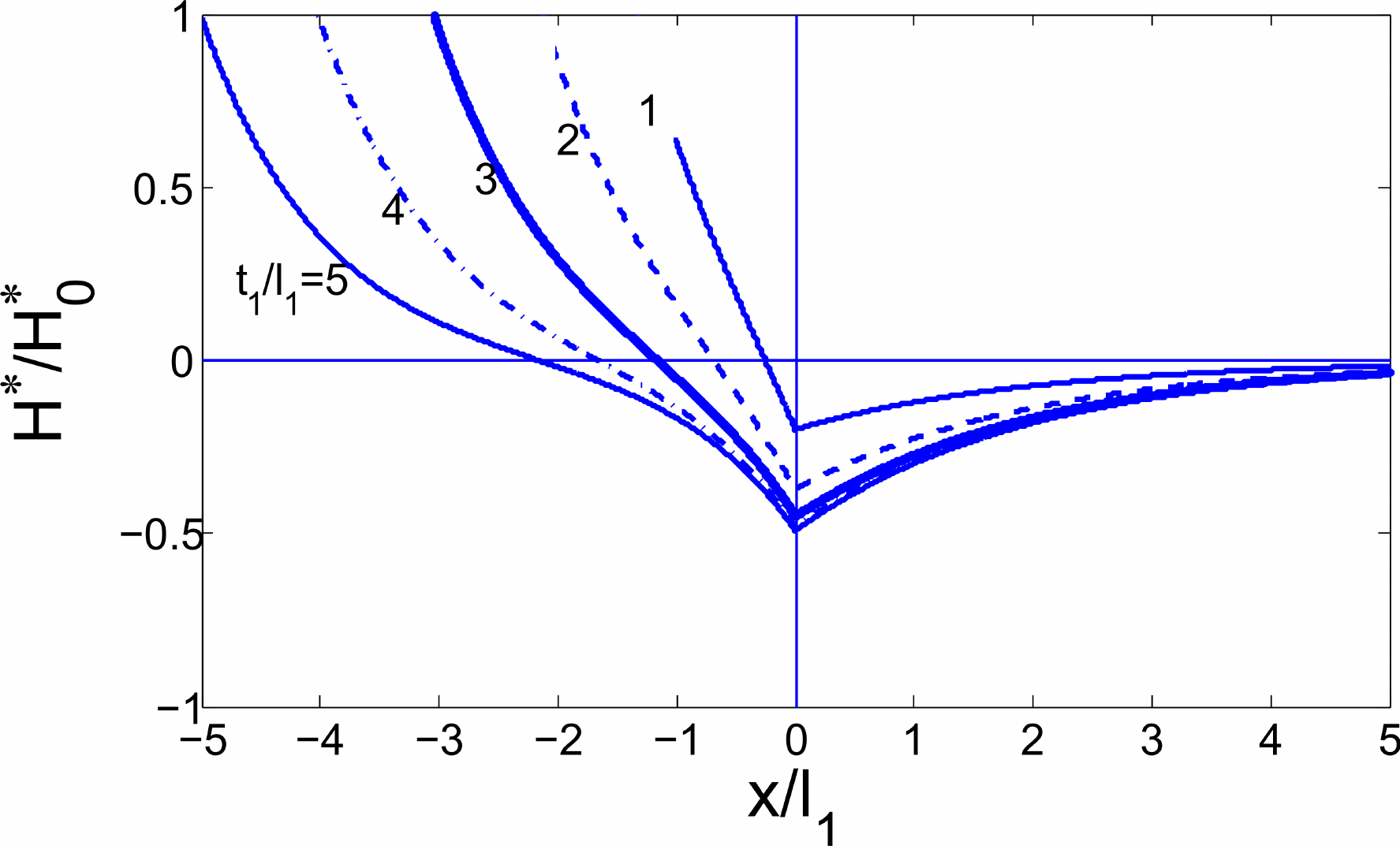}
\caption{Magnetization potential profiles $H^*$ for the bilayer of Fig.\ref{FIG:junction1}. Curves are from Eqs.(\ref{EQ:AC_h1}), (\ref{EQ:AC_h2}) and normalized to $H_0^{\ast}=j_{MS}/(l_2/\tau_2)$. Parameters are the same of Fig.\ref{FIG:junction1}} 
\label{FIG:junction2}
\end{figure}

\noindent By setting the boundary condition at the interface between the two media $H^{\ast}_1(0)=H^{\ast}_2(0)$ we find the value of the current at the interface

\begin{equation}
j_0=j_{MS}\frac{\cosh(t_1/l_1)-1}{\cosh(t_1/l_1)+r_{12}\sinh(t_1/l_1)\coth(t_2/l_2)}
\label{EQ:AC_j0}
\end{equation}

\noindent where $r_{12}=(l_1/\tau_1)/(l_2/\tau_2)$. Figs.\ref{FIG:junction1} and \ref{FIG:junction2} shows the profiles of the magnetization current and the effective field along the material for different values of $t_1/l_1$. 

\bibliography{00_SS}

\end{document}